\documentclass[12pt]{article}
\usepackage{graphicx,amsmath,epsfig}

\newlength{\dinwidth}
\newlength{\dinmargin}
\def\lapproxeq{\lower .7ex\hbox{$\;\stackrel{\textstyle                                                    
<}{\sim}\;$}}                                                    
\def\gapproxeq{\lower .7ex\hbox{$\;\stackrel{\textstyle                                                    
>}{\sim}\;$}}                                                    
\def\be{\begin{equation}}                                                    
\def\ee{\end{equation}}                                                    
\def\bea{\begin{eqnarray}}                                                    
\def\eea{\end{eqnarray}}

\def\sh{\hat s}
\def\sh2{{\hat s}^2}
\def\mbf#1{\mathchoice{\hbox{\boldmath $\displaystyle #1$}}
{\hbox{\boldmath $\textstyle #1$}}
{\hbox{\boldmath $\scriptstyle #1$}}
{\hbox{\boldmath $\scriptscriptstyle #1$}}}
\begin{document}                                                    
\titlepage                                                    
\begin{flushright}                                                    
IPPP/10/20  \\
DCPT/10/40 \\
MPP-2010-29\\                                                    
%\today \\                                                    
\end{flushright}                                                    
                                                    
\vspace*{0.5cm}                                                    
                                                    
\begin{center}                                                    
{\Large \bf Limiting soft particle emission 
in $e^+e^-$, hadronic and\\}
\vspace*{0.2cm}
{\Large \bf nuclear collisions}                                                                                                        
                                                    
\vspace*{1cm}                                                    
Wolfgang Ochs$^{a}$,Valery A. Khoze $^{b,c}$ and M.G. Ryskin$^{c}$\\                                                    
                                                   
\vspace*{0.5cm}  
$^a$ Max Planck Institut f\"ur Physik, Werner-Heisenberg-Institut,
F\"ohringer Ring 6,\\
 D-80805 Munich, Germany\\                                                 
$^b$ Institute for Particle Physics Phenomenology, University of Durham, DH1 3LE \\                                                   
$^c$ Petersburg Nuclear Physics Institute, Gatchina, St.~Petersburg, 188300, Russia\\           
                                                    
\end{center}                                                    
\vspace*{1cm}                                                    
                                                    
\begin{abstract}
In  $e^+e^-$ collisions the particle spectra at low momenta reflect the
properties of the underlying ``soft'' QCD gluon bremsstrahlung:
the particle density, in the limit
$p\to 0$, becomes
independent of
the incoming energy $\sqrt{s}$
and directly proportional
to the colour factors $C_A,C_F$ for primary gluons or
quarks respectively.
We find that experimental data from the
$pp$ and nuclear reactions reveal the same behaviour:
in the limit $p_T\to 0$
the invariant particle spectra become
independent of the collision energy,
and their intensities in
$e^+e^-$, $pp$ and nuclear reactions
are compatible with the expected
colour factors $C_F$: $ C_A$: $(N_{part}/2) C_A$  for $N_{part}$
nucleons, participating in the interaction.
Coherent
soft gluon bremsstrahlung is, therefore, suggested to be
the dominant QCD mechanism
for the soft particle production in all these reactions. These
``soft'' particles probe
the very early stage of hadron formation in the collision.
Future measurements at the LHC will provide crucial tests on the
contributions from possible incoherent multi-component processes.
\end{abstract}    
%\newpage           

\section{Introduction}

One of the interesting  phenomena in high energy collisions is the
production of {\it soft} particles with low momenta.
It is usually described in the existing  
%parton
  Monte Carlo models 
by a non-perturbative mechanism which transforms the  
 partonic final state
into a corresponding final state of hadrons. Best understood is the $e^+e^-$
process where the primarily produced $q\bar q$ pair evolves into a partonic
cascade and finally into hadrons. 
 A similar description applies to the
hadronic collisions with the high $p_T$ jets, where the partonic cascades are initiated 
by the { \it active} partons and the parton
{\it spectators}; 
%originally carried with the projectiles;
 also an incoherent
superposition of such processes could become important at high energies
in the TeV region. In nuclear collisions the collective phenomena play an 
important role as well. 

This paper is focused on the universal feature of the soft ($p\to 0$) 
particle production in all these collision processes which is based 
on the universality of the soft gluon bremsstrahlung  with the intensity
depending on the directions and colour charges of the participating partons 
but not on the collision
energy.
% in the soft limit $p\to 0$. 
From this point of view the perturbative
analysis of the quark gluon cascade (see, for example, 
\cite{Dokshitzer:1991wu,Khoze:1996dn})
is naturally combined with the idea of a close similarity between the parton and hadron spectra    
(``Local Parton Hadron Duality'' - LPHD \cite{Azimov:1984np}).
Such a scenario is deeply rooted in the space-time picture of 
the parton-cascade development and hadronization, see \cite{Dokshitzer:1991wu, Khoze:2000iq}.
This approach has been developed and tested in  details 
for the hadronic final states in $e^+e^-$ annihilation 
%(and partly in DIS)
 verifying, in
particular, the energy independence of the soft limit and the expected
dependence on the primary colour charges (quark or gluon jets), but it
should also be applicable 
 to other hard processes including $ep$ and $pp$ $(p\bar p)$ collisions
with high $p_T$ particle production, where the soft particles belong to the
% build the
%to the so-called
``underlying event'' \cite{Khoze:1996ij}~-~\cite{Khoze:1996ih}. A distinct dependence
on the primary colour charges makes this approach quite 
different from the simple
process-independent universality, for example, of a thermodynamical origin.  

Here, we are extending this approach to
the description of the so-called ``minimum bias'' events
in the
%to the untriggered
 hadronic and nuclear collisions
 where we expect similar effects generated by the
primary soft gluon  bremsstrahlung.
Note, first, that it is plausible to
expect that at high energies the bulk of the soft particles
originates in a semihard process from the fragmentation of the so-called
 minijets, which are produced by    
 the perturbative
mechanism similar to the $e^+e^-$ case.
 Next, recall that the perturbative
dynamics  results in the limiting behaviour of the soft particle densities, since the 
radiation of soft gluons from several quarks and gluons develops in a coherent
way: 
 different sources of secondaries should act as a single source with an
effective colour 
%(chiral/isotopic?) 
`charge' 
equal to the {\it vector} sum of
 all the colour `charges', 
since a gluon with a
large wave-length cannot resolve the smaller details 
(individual elements of the `colour antenna').
% smaller detail. 
It is worth emphasizing also that the low momentum hadrons are formed at the initial stage,
before the formation of fast particles \cite{Dokshitzer:1991wu, Khoze:2000iq}.
 Thus, the low-$p$ hadrons
emitted at small rapidities in the centre-of-mass frame ($c.m.s.$) 
bear no information
%'know' nothing
about the complicated structure of the whole event, in particular, about the 
secondaries in the fragmentation regions and the properties of the
initial hadrons.

It is natural to expect the  analogous interference phenomena in the
light hadron (pion) emission, when the particle inverse momentum, $1/p$,
which is a characteristics  of the resolution, exceeds the size of the
interaction region.  Thus, 
 we should observe the  limiting behaviour
in the soft particle densities in the proton-proton, proton-nuclear
 and nuclear-nuclear collisions
similar to that observed in the $e^+e^-$ annihilation.
All these processes can be characterized by 
%one can introduce
%\cite{Khoze:1996ij}
 the limiting density $I_0$ of the soft particle spectra \cite{Khoze:1996ij}
at vanishing rapidity $y$ and $p_T$ \footnote{We refer 
here to the one-side limit
$y\to 0$, not to the double-side limit $|y|\to 0$}
\begin{equation}
I_0=\left.\frac{1}{\sigma_{norm}}E\frac{d\sigma}{d^3p}\right|_{y\to0,p_T\to0}.
\label{invcs} 
\end{equation} 
In the case of  $e^+e^-$ annihilation
the normalization cross section is
$\sigma_{tot}$, while for the $pp$ and nuclear collisions
the rates  are often normalized by
$\sigma_{in}$=
$\sigma_{tot}-\sigma_{el}$. However, for the studies
of soft particle production 
it is more appropriate to use for normalization
another quantity, namely the %\mpar{***} 
so-called non-diffractive cross section, $\sigma_{ND}$,
defined as
$\sigma_{ND}$= $\sigma_{in}-\sigma_{D}$.
Here $\sigma_{D}$ describes the corresponding contribution
from the single diffractive and double diffractive
dissociation.
Strictly speaking, $\sigma_{D}$
depends on the collision energy (see, for example, a compilation
in Ref.\cite{dg} and the analysis in \cite{nns}).
However, this energy variation is
 within
the uncertainties of the current data, and 
it is plausible to evaluate the
non-diffractive cross section $\sigma_{ND}$
as being
10\% lower than $\sigma_{in}$.
%=\sigma_{tot}-\sigma_{el}$.
Note also, that here and in what follows the
quantity $I_0$ is defined as the density
averaged over the charged \footnote{In some cases, when we refer to the charged
pions, the value of $I_0$ is 7\% lower.}
particles %\mpar{****}
[$\frac{1}{2}(h^+ + h^-$)].
%\footnote{In some cases, when we refer to the charged
%pions, the value of $I_0$ is 7\% lower.}.
%
After recalling the relevant observations for
the $e^+e^-$ collisions we will turn to the
study of the limiting behaviour of
the density (\ref{invcs}) in the $pp$ and $AA$ collisions. 
We assume that these processes are initiated by the gluon exchanges, but more
general configurations are possible. 

The relation between parton and hadron final states is formulated
at first for inclusive observables such as we discuss here. It is an
interesting question how such relations can be realised by the
exclusive final states including resonances. In the limit $p\to 0$
soft gluons in the cascade add coherently and are represented by the Born
term; likewise the soft pions from a general final state should add
coherently to yield the energy independent density.

Analogies between the bremsstrahlung in QED and multiple particle 
production in
hadronic collisions date back to the pre-QCD times \cite{Feynman:1973xc,stodolsky}
and has lead in particular to the concept of {\it flat rapidity plateau}. 
Within QCD such an analogy was further developed
\cite{niedermayer} and the picture with one-gluon exchange
initiating high energy $pp$ collisions has appeared \cite{Low}. The present
QCD-based approach relates hadrons to the full quark-gluon cascade which
generates  particle jets, and the ``classical'' bremsstrahlung Born term is
dominant only in the ultrasoft limit $p_T\to 0$.

\section{Lessons from the soft particle spectra in $e^+e^-$ collisions}

Let us start from recalling the behaviour of the momentum spectrum of particles  
in $e^+e^-$ annihilation in the soft limit as the benchmark reaction. 
In this process it is convenient to measure
the inclusive momentum spectrum  since 
this does not require a determination of the jet axis 
which is needed for the spectra defined in terms of $y$ and $p_T$.

The existing data can be well described within 
the Modified Leading Logarithmic Approximation (MLLA) for the QCD parton
cascade together with LPHD for the description of hadrons 
(see, for instance \cite{Dokshitzer:1991wu,Khoze:1996dn}). For the very low
momenta a model dependent 
prescription is applied allowing to account for the mass effects. 
The perturbative expansion of the momentum spectrum of partons 
can be derived by an iteration of the Born term in 
the respective evolution equation. The
first %\mpar{****} 
term
%two terms
 in MLLA 
%have 
has the form \cite{Khoze:1996ij,Khoze:2000iq}
\begin{equation}
\frac{dn^{A,F}_g}{d\bar ydk_T^2}\ \simeq \ C_{A,F} \frac{\alpha_s(k_T)}{k_T^2}
%       \left( 1+{\cal O}\left(
%\ln\frac{\ln (p_T/\Lambda)}{\ln(Q_0/\Lambda)}\
%\ln\frac{\ln(p_T/(x\Lambda))}{\ln(p_T/\Lambda)}\right)\right)
\label{Born}               
\end{equation} 
%for
while the higher $\alpha_s$ corrections, which depend on the primary energy,
vanish as $\ln(\ln (k_T/\Lambda)/$ $\ln(Q_0/\Lambda))$ when the transverse
momentum of the gluon, $k_T$, aproaches its minimum (cut-off) value $Q_0$.
Here $n^{A,F}_g$ is the multiplicity of gluons emitted by the configuration
with the gluon ($A$) or quark ($F$) colour charge; $\bar y=\ln (1/x)$,
momentum fraction carried by the gluon is denoted as $x$ and QCD scale - as
$\Lambda$.
% In the
%perturbative calculation with massless gluons the soft singularity is regularized by the cut-off
%$p_T\ge Q_0$. The second term depends on 
%the primary energy
%through $\ln (1/x)$ at fixed momentum and vanishes if $p_T$
%approaches its minimum value $Q_0$ and the same is true for all higher order
%terms. 
In this limit the Born term remains and it is
independent of primary energy and directly proportional to the colour
factors. These features stay unchanged if the spectra are slightly modified
at low momentum within LPHD to include particles with mass and one arrives
at the perturbative expectation for the limiting behaviour
 \cite{Khoze:1996ij,Khoze:1997zq}
\begin{equation}
p\to 0:\qquad E\frac{dn}{d^3p}\to const.
\label{constlim}
\end{equation}
In this limit the gluon has a large wave-length and only resolves the
primarily produced partons, a radiation process represented by the Born term.
Another feature of the approximation (\ref{Born})
is the ``flat rapidity plateau'', that is,  the independence at fixed $k_T$
of the particle density on $\bar y$ or angular rapidity $y$.  This
approximation is true as far as we neglect the kinematical (threshold)
effects, when at large $y$ the energy of the observed secondary particle
becomes comparable with the energy of the initial interaction.

The available data show, that 
at the lowest accessible particle momenta, around $p=0.2$~GeV, the 
invariant density $E\frac{dn}{d^3p}$ 
for pions rises only by about 20\% over the $c.m.s.$ energy range 
$\sqrt{s}=3.0\div160 $ GeV, and with increasing momentum $p$ this rise becomes
stronger, see 
%\cite{Khoze:1996ij,Khoze:1997zq,Khoze:1996ih} 
\cite{Khoze:1996ij}~-~\cite{Khoze:1996ih}.
The MLLA formulae provide a rather good fit of the total energy and momentum
dependence of the observed one-particle spectrum, 
and clearly demonstrate the approach towards energy independence
of the spectrum as in the limit (\ref{constlim}).
Therefore, the observed rise of the particle density 
at mid-rapidities and the rise of the overall
multiplicity with energy increasing is caused by the high $p_T$ particles.
Note that in the soft limit $p\to 0$ one obtains
\begin{equation}
\left.E\frac{dn}{d^3p}\right|_{p\to 0}=
\left.E\frac{dn}{d^3p}\right|_{y\to 0,p_T\to 0},
 \label{I0+-1}
\end{equation}
since the invariant particle distribution approaches a constant value.
%\mpar{***}
% in this
%limiting case, and, therefore,
% it does not depend on the shape of the small phase space
%volume. 

The second important lesson from the $e^+e^-$ data 
concerns the dependence of the soft
particle density on the primary colour charges, that is,  on the difference
between the primary quark-antiquark and the gluon-gluon colour antennae-dipoles.
This was confirmed by the study of  3-jet ($q\bar q g$) events at different inter-jet angles.
In the limit when the gluon is parallel to a quark (antiquark) the
soft radiation pattern
is the same as that in the case of a $q\bar q$ dipole, since the soft gluon
cannot resolve the parallel quark and gluon. On the other hand, if the
$q\bar q$ system recoils against the gluon, it is seen by the emitted soft gluon as a
colour octet source. Then the soft radiation 
density in these two limiting cases should differ by the ratio of
colour factors
\begin{equation}
p\to 0:\qquad \left.E\frac{dn^{gg}}{d^3p}\right/
E\frac{dn^{q\bar q}}{d^3p}= C_A/C_F
 \label{ggtoqq}
\end{equation}
with  $C_A/C_F=9/4$, in agreement with (\ref{Born}). 
Experimentally, one cannot reach these extreme limits but we can derive
predictions on how the soft radiation density perpendicular to the 
event plane in the 3-jet event varies with the inter-jet 
angles between the two extremes
\cite{Khoze:1996ij}. These expectations are well confirmed by the DELPHI data
 \cite{DELPHI} which successfully reproduce the ratio (\ref{ggtoqq}). 
Note that the corresponding ratio for the global quark and gluon jet
multiplicities strongly deviates from (\ref{ggtoqq}) because of the
influence of the particles with higher $p_T$ and the importance of higher
orders in the perturbative MLLA calculation.

Neglecting the effects of order of $1/N_C^2$, the primary $q \bar q g$ antenna
can be represented by the two $q\bar q$ dipoles, where the gluon is replaced by the
parallel $q$ and $\bar q$.
In this approximation the ratio (\ref{ggtoqq})
becomes $C_A/C_F\to 2$, corresponding to the number of radiating dipoles. 
The radiation of soft gluons perpendicular to the
dipole varies with the opening angle 
$\Theta$ as $E\frac{dn}{d^3p}\sim(1-\cos \Theta)$. 
The experimental results demonstrate that the soft particle density follows
the lowest order QCD expectations for the given configuration of colour
charges, see for a review \cite{Khoze:2000iq}.

\section{Soft particle spectra in high energy $pp$ $(p\bar p)$ collisions}

\subsection{Minimal model for the minimum bias events}
In analogy to the $e^+e^-$ process, let us start from the minimal partonic process,
which can be responsible for the very soft gluon bremsstrahlung.
We assume here that the underlying physics of the minimum bias events
is based on the collisions of two partons within the protons. 
The partonic process of lowest perturbative 
order corresponds to  one gluon exchange, which leads to a
dominantly small scattering angle and a non-vanishing cross section 
at high energies. In the case of  elastic scattering between 
the two incoming
quarks the exchange of the $t$-channel gluon rearranges 
%annihilates 
the incoming colours and creates
the outgoing colour charges which leads to the radiation of the soft gluons
from the effective colour octet dipole.
For a large number of colours in the above-mentioned approximation
 this radiation
can be described as being generated by a superposition
of the { \it two } aligned dipoles.
%\
In the case of small-angle scattering
the same soft gluon radiation pattern  appears also with %\mpar{***}
the incoming quarks replaced by 
%one or two 
gluons, as can be explicitly
seen from the radiation patterns given in \cite{EKS}.

Therefore, the limiting soft radiation densities in the $pp$
and $e^+e^-$ collisions should differ by a factor of 
\begin{equation}
p\to 0:\qquad   I_0^{pp}/I_0^{e^+e^-}\approx C_A/C_F \label{ratio94}
\end{equation}   
at all energies, as in the ratio in Eq. (\ref{ggtoqq})
for the spectra induced by the $gg$ 
and $q\bar q$ dipoles. 

For a similar reason the particle 
multiplicity ratio for the two processes is expected 
to be equal to the same quantity 
 $C_A/C_F$  \cite{bg}.
In our case, we anticipate that this
QCD expectation could be valid only in a specific soft limiting
case, while
as we discuss, the $p_T$-behaviour 
of the spectra in both processes are different
and energy dependent, therefore the total multiplicities should differ 
as well.

One can also consider more complex situations corresponding to the
 multiple gluon exchange. However,
if there is a multiple gluon exchange between the pair of 
quarks, then the total colour of
the exchanged system could be only an octet or a singlet; 
since the singlet exchange rather corresponds
 to a diffractive process, its contribution to 
the central production is
negligible,   and we come back to the previous case. 
If the multiple gluon exchange involves both quarks and
 spectator diquarks, but does not destroy the diquark,
(which acts as a local object), then the colour of the
 'quark-diquark' system can be either
an octet or a singlet,
 and, thus, again after the multiple gluon exchange this
generates in the $t$-channel only the 'octet' colour flow.
Finally, recall that 
in the case of 
more complicated diagrams the dominant
Leading Logarithmic contribution in
both the DGLAP and BFKL evolutions also corresponds to the colour 
octet exchange.

In the early discussions \cite{early}
several specific models were proposed, which predicted 
the ratio of the central rapidity particle densities in the
 $pp$ and $e^+e^-$ collisions.
In these models the
final state in $pp$-interactions is constructed from
the two chains of particles connecting $q$ and $qq$
while there is one chain between $q$ and $\bar q$ in  $e^+e^-$ collisions, 
but the ratio 
of these densities in both processes depends on the amount of overlap between 
both of the chains and can vary between one and two. Later on, 
the semi-hard gluon bremsstrahlung has
been included in the models and this leads to a rapidity plateau  rising with 
energy,
and a simple relation between both processes fades away.
An application of the Lund model to hadronic scattering at the energies
below ISR was discussed in \cite{gg1}, where the same one-string 
description was used for both, $e^+e^-$ and $pp$ collisions.

At higher energies this scenario should
be modified  in order to
incorporate the multiple interactions (MI) 
of partons (that is simultaneous interactions
of two or more pairs of partons) 
 and minijet formation, which
would lead to a certain energy growth of $I_0$ 
 \footnote{We thank Gosta Gustafson
for a  discussion of the Lund model results.},
for recent development  based on the dipole cascades and multiple collisions,
see \cite{gg2}. 

Current Monte Carlo models include the MI option, which
in terms of a simple eikonal model for soft $pp$-scattering,
where one interaction  corresponds to the exchange of a cut {\it bare}
Pomeron, is described by the contribution of the few cut Pomerons.
 Particle density produced in the central region by one cut Pomeron is
energy independent
\footnote{In a simple soft
scattering model particle density is energy-independent.
%similarly to string fragmentation mechanism.  
For shower MCs, where each MI is modeled by the 
perturbative QCD contribution of the form of
$\int dx_1dx_2 g(x_1,q_t)g(x_2,q_t)d\hat\sigma/ d^2q_t|_{q_t>q_{min}}$,
the approximately constant behaviour results from a
 compensation between the growth of the gluon density 
$xg(x,q_t)\sim x^{-\lambda}$ with decreasing momentum fraction $x\sim q_t/\sqrt{s}$ and an increase
of a cutoff
 $q_{min}\sim 1/s^{0.08\div 0.13}$.}.
Recall, that the cross
section of hard subprocess is $d\hat\sigma/d^2q_t\propto 1/q^4_t$.
However, in  the Gribov-Regge theory
the mean number of MIs, that is the number of cut Pomerons, increases
as $s^\Delta$, where the bare Pomeron intercept is
$\alpha_P(0)=1+\Delta$
(in terms of  the gluon density $xg(x,q_t)\sim x^{-\lambda}$ we  identify
$\Delta=\lambda$) \footnote{In the recent model for soft  hadronic
interactions \cite{nns}
 $\Delta \sim 0.3 $, which is in agreement with the resummed NLL BFKL result,
for example \cite{spread}.}.
This is the well known AGK result \cite{AGK} -- single
particle inclusive cross section is described
 just by  one Pomeron exchange, and,  thanks to the AGK cancellation,
there is no absorptive corrections.
Therefore, we arrive at
$d\sigma/d^3p\propto s^\Delta$, while the growth of inelastic cross section
is reduced by the absorptive effects (asymptotically instead of
$\sigma_{in}\propto s^\Delta$ we should reach the Froisart limit
 $\sigma_{in}\propto \ln^2 s$). Phenomenologically, the energy behaviour of
$\sigma_{in}$ is close to $s^\epsilon$ with $\epsilon\sim 0.08\div 0.1$.
Thus, the particle density $I_0=(1/\sigma_{in}) Ed\sigma/d^3p\propto
s^{\Delta-\epsilon}$,
 and this growth with energy reflects
 the increasing number of MIs, which act in the present Monte Carlo models
incoherently.
An account of the coherence effect should diminish the multiplicity of 
low $p_T$-secondaries produced by MI.
An explicit computation \cite{TS2} using Perugia P0 tune \cite{TS1}
of the Pythia MC model, shows that in the $c.m.s.$ energy interval
$\sqrt{s}=23\div14000 $ GeV the invariant density
in the $pp$ collisions $E\frac{dn}{d^3p}$ 
in the range $p_T = 0\div 0.5$ GeV and $y=0\div 0.5$ 
is rising by a factor of about 2. 
Other, 'more aggressive', tunes may result in even higher 
rise, up to a factor of 3 at most \footnote{We are very grateful
to Torbjorn Sjostrand for the discussion
of this and other issues related to results
in Pythia model.}.
As we discussed above,
without incorporation of coherence, such increase should hold
on up to the asymptotic energies.

Finally, from such simple properties as eq. (\ref{ratio94}) one should not
conclude that these two processes are intrinsically 
very similar beyond the considered
limit $p_T\to 0$ and, indeed, the above models 
show very different phenomena. 
In fact, the event structure of a 2-jet system in $e^+e^-$
annihilation is not the same as that of a system with two (q-qq) pairs in $pp$
collisions with its much larger fluctuation properties. However, in the
considered limit, the produced partons (q-qq) are resolved as an extended
colour octet source with the enhanced radiation according to eq.
(\ref{ratio94}). Such intuitive limits may help to understand the low $p_T$
production based on a perturbative ansatz.

\subsection{Energy dependence of the limiting soft spectrum}
For $pp$ interactions %\mpar{***}
in case of the minimal model discussed above the soft radiation
pattern should again be energy
independent analogously
 to the  $e^+e^-$ case, but with the intensity
about a factor of two higher.
We are analyzing below the $p_T$-spectra
in $pp$ collisions 
in order to determine the quantity $I_0$ 
for $p_T\to 0$ at the central rapidity $y=0$, as defined by relation
(\ref{invcs}).
The previous analysis of the invariant cross section
$\frac{E}{\sigma_{in}}\frac{d\sigma}{d^3p}$ 
in the $p_T$-range $0.3 \div 1.0$ GeV indicated 
 a moderate rise with energy increasing from the ISR %\cite{Breakstone:1995ug}
to the S$p\bar p$S colliders, % \cite{Albajar:1989an}, 
see Ref. \cite{Khoze:1996ih}. %Nijmegen report.  
Later on, further data from  CDF at the
Tevatron at higher energies up to 1800 GeV \cite{Abe:1988yu} and from
RHIC became available, which required a fresh look at this problem.

The inclusive spectra were measured down to the transverse momenta of
$0.1\div 0.4$~GeV. Some 
experimental groups  performed fits using certain simple parametrizations
of the spectra from which 
the quantity $I_0$ can be  easily determined.
The British Scandinavian Collaboration (BS)
%\cite{Alper:1974rw} only high pt > 2 GeV
\cite{Alper:1975jm}
fitted the invariant cross sections at different energies between
$\sqrt{s}=23$ to 63 GeV as
\begin{equation}
E\frac{d\sigma}{d^3p}=A \exp{(Bp_T+Cp_T^2+Dy^2)},
\label{bsparam}
\end{equation}
 while
the  UA1 \cite{Albajar:1989an}
and CDF \cite{Abe:1988yu} groups used the parametrization 
\begin{equation}
E\frac{d\sigma}{d^3p}=A(1+p_T/p_0)^{-n}. \label{ua1param} 
\end{equation}
From these fits 
the quantity (\ref{invcs}) can be
found as $I_0=A/\sigma_{in}$. The BS collaboration \cite{Alper:1975jm}
confirmed the exponential behaviour down to the pion momenta as low as $p_T\sim
0.1$~GeV.
Correspondingly, we derived the quantity $I_0$ from the STAR data \cite{star1}  
using the exponential extrapolation 
of their charged particle spectra. 

As the results depend on the function chosen for extrapolation we also
considered other functional forms. 
The PHOBOS~\cite{phobos} collaboration has measured the spectra down to very
small $p_T=0.03$ GeV in $AuAu$ collisions at 200 GeV. These data 
indicate a flattening of the distribution towards small $p_T$ according
to a functional form following from a thermal model
\begin{equation}
E\frac{dn}{d^3p}=\frac{A}{\exp(m_T/T)-1}
\label{thermal}
\end{equation}
with $m_T=\sqrt{m^2+p_T^2}$ and temperature $T$ as parameter.

In order to test the dependence of the extrapolated results on the choice of
the function we compared the $pp$ data of highest accuracy which is provided by
the STAR collaboration \cite{star1} with errors of $\pm3$\% to an
exponential as in (\ref{bsparam}) (but with $C=0$) and the function
(\ref{thermal}) in the $p_T$ ranges 0.225-0.425 GeV and 0.225-0.625 GeV.
We found that the extrapolated cross section at $p_T=0$ is reduced by  
25\% for the thermal fit.

It should be noted that the forms (\ref{bsparam}) and (\ref{ua1param})
do not have the correct analytic behaviour
at $p_T\to 0$ where they behave linearly in $p_T$ and the invariant 
cross section is not regular; 
therefore, one expects to obtain an overestimate from such fits.
This problem does not occur with the non-singular fits of the
type (\ref{thermal}) with quadratic behaviour in $p_T$.

Experimental results over a range of energies are only available
from the exponential and power like fits which we consider next.
%First we consider the published results from the exponential fits.
To obtain the  cross section, $\sigma_{in}$,
we take the total and elastic cross section data for the $pp$ (BS) and
$p\bar p$
 scattering (UA1, CDF), as collected by the Particle Data
Group \cite{pdg}. The results are shown in Table 1 after an
interpolation.
In order to determine the quantity $I_0$ for the STAR entry in this Table,
we renormalized their minimum bias data to the $\sigma_{in}$ as well.
%Here and in what follows
%we take 
%the non-diffractive cross section to be 
%10\% lower than the inelastic cross section,
%as can be concluded from the data compilation given in Ref. \cite{dg}
%and the analysis performed in \cite{nns}.
%\mpar{Mod}
\begin{table}[t]
\caption{Soft limit $I_0=A/\sigma_{in}$ from the exponential fits to the
single charged-particle 
$[(h^++h^-)/2]$ spectra in the $pp\ (p\bar p)$ collisions; for the BS and STAR data we sum over the 
$\pi^\pm,K^\pm,p^\pm$  distributions.} 
%$I_0$ for STAR from the
%exponential extrapolation of
%the charged-particle spectra, but the normalisation cross-section
%is reduced in order to account for
%the diffractive dissociation.}

\begin{tabular}{lcccccccc}
\hline
Exp &$\sqrt{s}$    & $p_{T,min}$ & A  &   $\sigma_{tot}$ & $\sigma_{el}$
    &  $\sigma_{in}$ & $I_0=A/\sigma_{in}$\\
    &[GeV] & [GeV] & [mb/GeV$^2$] & [mb] & [mb] & [mb] & [GeV$^{-2)}$]\\
\hline
BS & 23 & 0.1 & $191\pm7$ & 39.4 &6.8 & $32.6\pm 0.5$ &$ 5.9\pm 0.3$ \\ 
BS & 45 & 0.3 & $238 \pm 7$ & 41.9 &7.5 & $34.4\pm0.7 $ &$ 6.9\pm 0.3$ \\
BS & 63 & 0.1& $307\pm 20 $ & 43.0 &7.8 & $35.2\pm0.6$ & $ 8.7\pm0.7$ \\ 
STAR &200 & 0.2&            &    &  & & $7.5\pm 0.8$ \\   
UA1 & 200 &0.25 & $286\pm 17$ & 52 & 9.2 &$ 43\pm4$ & $6.6\pm0.7$\\ 
UA1 & 500 & 0.25& $408\pm24$ & 62 & 13 & $49\pm2$ & $8.3\pm0.6$\\
CDF & 630 & 0.4 & $300\pm 20$ & 63 & 13 & $50\pm2$ &$ 6.0\pm 0.5$\\
UA1 & 900 & 0.25& $382\pm 20$ & 68 & 15 & $53\pm 4$ & $7.2\pm0.7$\\
CDF & 1800 & 0.4 & $450\pm 10$ & 74 & 17 & $57\pm3$ & $7.9\pm 0.5$\\
\hline
\label{tab:i0}
\end{tabular}
\end{table}

\begin{figure}[h]
\begin{center}
\mbox{\epsfig{file=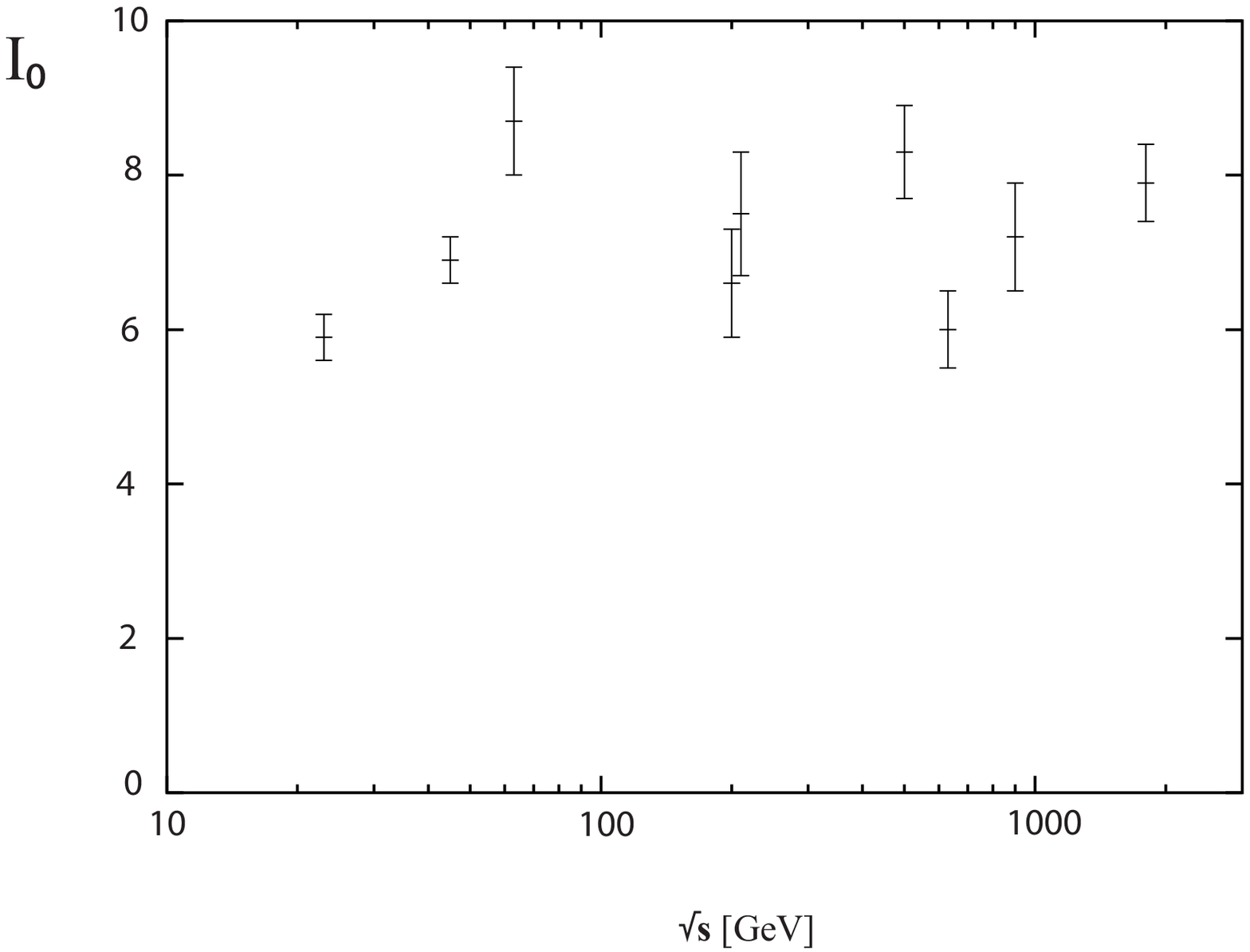,angle=0,bb=0 30 510 380,clip=,width=12cm}}
%bbllx=0cm,bblly=4.0cm,%
%bburx=22.0cm,bbury=18.cm,width=12.cm}}\\
%\mbox{\epsfig{file=i0X2.eps,angle=0,width=12.cm}}
\end{center}
\vspace{-0.5cm}
\hspace*{12cm}{\bf ${\bf \sqrt{s}}$ [GeV] }
\caption{Soft limit $I_0$ of the invariant density $E\frac{dn}{d^3p}$ 
of charged particles [$(h^+ +h^-)/2$] in $pp$ collisions as a function of
c.m.s. energy $\sqrt{s}$ (from exponential extrapolation, Table 1).}
\label{fig:I0}
\end{figure}
The data for $I_0$ in the last column of Table 1, shown in Fig. \ref{fig:I0}, indicate an initial rise in
the BS data with
increasing energy, but it does not hold at higher energies. Remarkably, the
data in the large energy range of $40\div 1800$ GeV 
show a rather energy-independent behaviour, and they  
fluctuate around some common mean value by about $\pm15$\%:
% \pm 15\%:  
%
\begin{equation}
{\rm inelastic }\ pp/p\bar p\ {\rm collisions\ (exp.\ fit):}\quad 
 I_0\approx (7\pm 1)\ \text{GeV}^{-2}
\text{          }\label{I0pp}
\end{equation}
and for minimum bias events with the above estimate for diffractive cross
sections
\begin{equation}
{\rm non-diffractive} \ pp/p\bar p\ {\rm collisions\ (exp.\ fit):}\ I_0\approx (8\pm 1)\
\text{GeV}^{-2}
\text{          }\label{I0ppnd}
\end{equation}
Next we consider the extrapolation with the thermal function
(\ref{thermal}).
While our main finding, the lack of any major energy dependence of $I_0$ does
not depend on the extrapolation procedure the central value of $I_0$ will be
reduced by 25\% and we obtain 
\begin{equation}
{\rm non-diffractive} \ pp/p\bar p\ {\rm collisions\ (therm.\ fit):}\
 I_0\approx (6\pm1)\ \text{GeV}^{-2}
\text{          }\label{I0ppndth}
\end{equation}

This observation of energy independence is  similar to that in $e^+e^-$ annihilation,
discussed above, and it is
consistent with the 
expectation  corresponding to
a single coherent bremsstrahlung
process.  On the other hand, for an incoherent superposition of many processes 
one would rather expect  $I_0$ values, which rise
%that the $I_0$ quantity
with energy increasing since the number of such processes 
would typically increase.
We should also recall that the very weak energy dependence (if any)
of the quantity $I_0$ contrasts to the energy dependence of the
$p_T$-integrated central rapidity density
$dn/d\eta|_{\eta=0}$. The latter observable is rising in the considered
energy range by a factor 2-3 in  the pp and AA collisions (see, for example,
the recent discussions in \cite{sarkisyan}).

Since the soft gluon emission is driven by the lowest order
diagram, according to eq.
(\ref{Born}),
we  may expect that
the same limiting behaviour ($d\sigma/dyd^2p_T\to const$ for $p_T\to 0$) should be observed
at any rapidity and not {\it only} at $y=0$ and we
expect a flat rapidity plateau in this limit, as mentioned above. 
This is true, as long as we can neglect the
kinematical (threshold) effects for large rapidity $y$.
Also, we are looking for the soft hadrons
(pions) which are formed first and do not participate in
 further collisions with the
faster secondaries. These conditions are better satisfied for the
particles in the central region. For our analysis we use the
inclusive cross sections at $y=0$ where we have the best collection of data.
Actually, the $y$-dependence was studied by the  BS collaboration [22] at
the ISR.
In the rapidity range $|y|<1.5$ and $p_T<1$ GeV  this dependence is rather weak
and the Gaussian fit as in eq. (7) yields values around $D\sim 0.1$. Yet
smaller values of $D$ are required if additional data from large $y$ are
included. Measurements at higher energies would be of interest as well.

%\mpar{par. added}
Finally, we comment on the first publication of charged particle $p_T$ spectra in
non-diffractive minimum bias events from the LHC at $\sqrt{s}=0.9$ and 
2.36 TeV by the CMS collaboration \cite{cms}. One can observe the
convergence of the spectra for $p_T\to 0$. An exponential 
extrapolation of the data to this limit yields $I_0\approx 9$ GeV$^{-2}$, %\mpar{***}
which is
consistent with (\ref{I0ppnd}). Further studies at higher energies at
LHC will provide a critical test of the contributions of additional
incoherent multiple interaction components.
%********************************************************************
%
\subsection{Comparison with the $e^+e^-$ data}
It is also instructive to compare the
% experimental
 observation (\ref{I0ppnd}) 
with the corresponding quantity for the $e^+e^-$ annihilation.
As we already discussed, 
since in the $pp$ collisions the diffractive events lead only to a small
contribution to the central rapidity region, the
diffractive piece should be subtracted in this comparison. 
%The non-diffractive cross section is by about 
%10\% lower than the inelastic cross section used for normalization in Tab.
%1, as can be concluded from the data compilation given in Ref. \cite{dg}
%and the analysis performed in \cite{nns}. 
We estimate the limiting value
$I_0^{e^+e^-}$ in three different ways:

1. The TPC/2$\gamma$ collaboration \cite{tpc} 
%has
 compared the results for inclusive 
 $\pi/K/p$ particle spectra, measured in the $e^+e^-$ annihilation at 29 GeV, 
with the corresponding spectra 
from the non-diffractive $pp$ collisions at 53 GeV, obtained by the BS collaboration
\cite{Alper:1975jm}.
The higher $pp$ collision energy was chosen in order to take into account the
lower effective energies of the parton-parton collisions.
 
The spectra measured in  both reactions have comparable magnitude in the
transverse momentum range $0.25<p_T<0.5$ GeV, where 
the rapidity $y$ and
$p_T$ in  $e^+e^-$ annihilation events are defined with
respect to the sphericity axis.
As discussed above, the $pp$ data \cite{Alper:1975jm} 
follow the exponential behaviour for small
$p_T< 1$~GeV down to $p_T\sim 0.1$ GeV, 
and one can extrapolate the data for pions ($\pi^+$ or $\pi^-$ 
reproduced in \cite{tpc})\footnote{The data plotted in Fig. 2 of
\cite{tpc} should be divided by $2\pi$ in order
to obtain the quantity $\frac{E}{\sigma}\frac{d\sigma}{d^3p}$.} to
$I_0\approx 8$ GeV$^{-2}$ in agreement with (\ref{I0ppnd}) or to a 25\%
lower value if we take into account mass effects as above.
On the other hand, the  $p_T$-spectra in $e^+e^-$ annihilation  
\cite{tpc} deviate from the exponential behaviour, decreasing more weakly for higher
$p_T$ and flattening  towards the small $p_T$-values with an extrapolated limit
$I_0\approx (3.0\pm 0.3)$ GeV$^{-2}$. This way
we arrive for the pions at the ratio
$r\equiv I_0^{pp}/ I_0^{e^+e^-}\approx 2.7$ or at $r\approx 2.0$ if we apply
again the 25\% reduction for mass effects in the $pp$ analysis.
For the kaons an exponential extrapolation appears to be applicable for both
processes at $p_T< 1$ GeV, but with different slopes, 
and the corresponding values are 
$I_0\approx 0.48$ GeV$^{-2}$ and $I_0\approx 0.24$ GeV$^{-2}$
so that  $I_0^{pp}/ I_0^{e^+e^-}\approx 2$ in the case of kaons as well.
While the pion and kaon spectra are steeper in $pp$ collisions and cross
those in $e^+e^-$ collisions,
the two spectra for protons do not show this behaviour but are rather
similar in the considered range. Then we look at other results as well.  

 \begin{figure}[t]
\begin{center}
%\mbox{\epsfig{file=ednd3p.ps,width=9.cm}}
\mbox{\epsfig{file=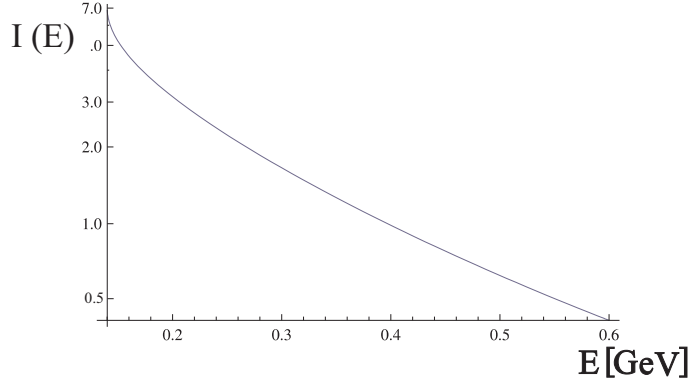,width=9.cm}}
\end{center}
\caption{The density
$I(E)=\frac{E}{4\pi p^2}\frac{dn}{dp}$ of pions
[$(\pi^+ +\pi^-)/2$] as a function of the $c.m.s.$ pion energy $E$, 
as derived from the fit to the BS $pp$
data and normalized to the non-diffractive cross section at 53 GeV.}
\label{fig:ednd3p}
 \end{figure}

2. Since no other group  presented their results in terms of the  $y,p_T$
variables in the $e^+e^-$ experiments,  we consider here also the 
extrapolation in the standard variables, momentum $p$ or the energy 
$E$, which are directly measurable without any reference to a jet axis. 
Then in the $e^+e^-$ collisions one considers the distribution  
$\frac{E}{4\pi p^2}\frac{dn}{dp}$.
%Assuming a distribution exponential in  $p_T$ with respect to a primary jet
%axis as in $pp$ collisions  \cite{Alper:1975jm}
%we can calculate the shape of the invariant spectrum 
%$\frac{E}{\sigma}\frac{d3\sigma}{d3p}$ in momentum $p$ or
%energy $E$.
Such spectrum can be also determined in the $pp$ collisions, and we calculate its
shape using the parametrisation 
for the   invariant spectrum of the 
pions [$(\pi^++\pi^-)/2$] at 53 GeV 
$
E\frac{d\sigma}{d^3p}=\frac{1}{\pi}\frac{d\sigma}{dydp_T^2}\equiv
I(y,p_T)\ \sigma, 
$
as given by the BS collaboration~\cite{Alper:1975jm}, see Eq. (\ref{bsparam}),
with the parameters
$A=212$ mb/GeV$^2$, $B=-7.3$ GeV$^{-1}$, $C=1.2$ GeV$^{-2}$ and
$D=-0.13$ and with normalization by the non-diffractive cross section
as in \cite{tpc}.
We calculate the corresponding distribution over the $c.m.s.$ momentum $p$,
using $dy=dp_z/E$ as
\begin{eqnarray}
\frac{dn}{dp}&=& 2 \int_0^p dp_z \int_0^{p^2}dp_T^2 \frac{dn}{dp_z dp_T^2}
\delta(p-\sqrt{p_z^2+p_T^2})\\ 
%       &=&\frac{4\pi p}{E(p)}  \int_0^p dp_z I(y(p,p_z),p_T(p,p_z))\\
\frac{E}{4\pi p^2}\frac{dn}{dp}
   &=&  \int_0^p \frac{dp_z}{p} I(y(p,p_z),p_T(p,p_z))
\label{ednd3p}
\end{eqnarray}
where $y=\frac{1}{2} \ln\frac{E(p)+p_z}{E(p)-p_z}$ and
$p_T=\sqrt{p^2-p_z^2}$. This spectrum 
is shown in  
Fig. \ref{fig:ednd3p} as a function of $c.m.s.$ pion energy $E$. 
One can see the peak structure at low energy $E$ \footnote{ 
The Gaussian parametrization of the distribution over the rapidity $y$
seems to be a bit  low at $y>1$ (see Fig. 24 of \cite{Alper:1975jm}),
but this  affects the spectrum shown in Fig. 2 by less than $\sim$ 5\%.}.
Pion distributions over the $c.m.s.$ energy, $E$, from the $e^+e^-$ annihilation 
are collected in Fig. 3a
of \cite{Khoze:1996ij}. The available data, especially ARGUS results
\cite{argus} at $\sqrt{s}\sim 10$
GeV, extending towards the small momenta $p\sim 0.05$ GeV, do not show
this kind of peak. They rather show a flatter exponential distribution
over $E$
with the limit $I_0\approx 3$ GeV$^{-2}$ for  $E\to m_\pi$, about half the
$pp$ value.
\footnote{The data plotted in  \cite{Khoze:1996ij}
refer to [$\pi^+ + \pi^-$], and we have to multiply 
the results by a factor 1/2.}
The $e^+e^-$ data at higher energies $E$ are rising with $\sqrt{s}$ and will
cross eventually the spectrum for $pp$ collisions in
Fig. \ref{fig:ednd3p}.
Note that the dependence on the jet-axis
definition disappears for $p\to 0$. These results demonstrate again that the
inclusive spectra in $e^+e^-$- are flatter than in $pp$ collisions at the
low values of $E$.    

3.  The TASSO collaboration \cite{tasso} has presented fits to the
invariant spectrum with exponential 
%terms 
 form in $E$ %\mpar{***}
% as well to the form
\begin{equation}
\frac{E}{4\pi p^2}\frac{d\sigma}{dp} = \sum_m A_m \exp(-B_mE),
\label{tassofit}
\end{equation}
with 2 or 3 terms,
which we can use to obtain $I_0$ from the limit $E\to m_\pi$.
For normalization we take the total cross section as
$\sigma_{tot}=\sigma_{\mu\mu}R$ with $R\approx 4$ representing the data
collection in \cite{pdg} in this energy region.

Results based on the TASSO fits and our estimates from the experiments at  lower $e^+e^-$
 energies with small momentum cut-off are summarized in Table 
\ref{tab:I0e+-}. They show the consistent values for $I_0$. At higher energies no fits 
allowing to perform the extrapolation
are available. However, as already noted in Sect. 2, the energy
dependence at $p=0.2$ GeV is weak over the energy range $\sqrt{s} \approx
3\div 160$ GeV,
and the deviations from the model incorporating QCD coherence are below $\sim$15\%.
Therefore, we evaluate $I_0$ from Table \ref{tab:I0e+-} as
\begin{equation}
e^+e^-\ {\rm annihilation:}\ I_0^{e^+e^-}\approx (3.3\pm 0.5)\ \text{GeV}^{-2}.
\text{          }\label{I0+-}
\end{equation}

The error in this result should include the uncertainty caused by
the extrapolation (\ref{tassofit}) over the energy $E$, which also
allows to fit the data at very low momenta $p\sim 0.05$ GeV.

\begin{table}[h]
\caption{Soft limit $I_0$ in the $e^+e^-$ collisions of the fits 
(\ref{tassofit}) (leading term $A_1,B_1$) from TASSO~\cite{tasso} to the
single pion energy spectra
$[(\pi^++\pi^-)/2]$ and the estimates, found  using data from ARGUS~\cite{argus} and
TPC/2$\gamma$ \cite{tpc}.}
\begin{tabular}{lccccccc}
\hline
Exp &$\sqrt{s}$    & $p_{min}$ & $A_1$  & $B_1$  & $\sigma_{tot}$
     & $I_0$\\
    &[GeV] & [GeV] & [nb/GeV$^2$] & [GeV$^{-1}$] & [nb] & [GeV$^{-2)}$]\\
\hline
ARGUS & 10 & $p>0.05$ &  &  &  &$ 3.0\pm 0.5$ \\ 
TASSO & 14 & $p>0.3$ & $23.9 \pm 3.9$ & $5.25\pm 0.43$ &1.77 & $ 3.5\pm 0.6$ \\
TASSO & 22 & $p>0.3$ & $8.0 \pm 1.2$ & $4.70\pm 0.32$ &0.72 & $ 3.4\pm 0.5$ \\
TASSO & 34 & $p>0.3$ & $3.7 \pm 0.6$ & $4.97\pm 0.45$ &0.30 & $ 3.4\pm 0.5$ \\
TPC/2$\gamma$ & 29 & $p_T\sim 0.05$ &  & & &$ 3.0\pm 0.3$ \\
\hline
\label{tab:I0e+-}
\end{tabular}
\end{table}

It is also interesting to compare with DIS results in the current
fragmentation region in  the Breit frame,
 which should correspond to the quark
fragmentation similar to $e^+e^-$ annihilation. 
 The H1 Collaboration at HERA
\cite{h1} performed a study of this type at $Q^2$ varying from 12 to
100 GeV$^2$.  Indeed, they found that the  particle density at low momenta
is nearly constant.
The soft limit is obtained by extrapolation of the data in
energy towards $E=Q_0=270$ MeV, which is used as an effective mass of 
charged particles. Extrapolating the data points at the lowest energy above
$E=0.3$ GeV to $E=Q_0$ at  $<Q>=19.6$ GeV we arrive at
$I_0\approx 4\pm 0.5$ GeV$^{-2}$, which is compatible with Eq. (\ref{I0+-}). It would
be interesting to analyse the recent more detailed data from ZEUS \cite{zeus} 
in a similar way.

Finally,
we can compare with the $pp$ collision data,
using the estimates (\ref{I0ppnd}) and  (\ref{I0ppndth}) 
for the non-diffractive events.
Then we obtain
\begin{equation}
I_0^{pp} / I_0^{e^+e^-}  \approx (1.8\pm 0.4) \div  (2.4\pm 0.5),
\text{          }\label{I0pe}  
\end{equation}
where the two numbers correspond to the thermal
or exponential parametrization respectively.
As discussed above, the evaluation  based on the exponential
fit may be considered as an upper limit, which would apply for a distribution
without accounting for a mass effect. The uncertainty in (\ref{I0pe}) will be largely reduced
by more precise data on the spectrum at  very small $p_T$ in $pp$
collisions.

The result (\ref{I0pe}) agrees well with the expectation
from the different primary quark and gluon sources in 
(\ref{ratio94}).

The approximate energy independence of the quantity $I_0$  in both
collision processes, as well as the obtained ratio of the $I_0$ values in the
 $pp$ and $e^+e^-$ reactions, which is about 2,
%within a factor of 2 
are remarkable. This is
a serious argument in favour 
of the relevance of the
elementary bremsstrahlung process, also for the soft $pp$ interactions.

\section{Nucleus-Nucleus interaction}
\subsection{Spectra at low transverse momenta}
In the case of $AA$ scattering one would naively expect,
that the soft particle density is equal to that in the $pp$-collisions
times the mean number of nucleon-nucleon collisions,~$N_{coll}$. 
Such an estimate should be valid for the point-like interactions,
 for example, for the
events with the large $p_T$ particle production, if we
neglect the energy losses  in the nuclear medium.
However, at low momentum transfer,
$p_T$, the coherence effects should reduce the
particle production, in particular,  due to the 
 destructive interference between the different
amplitudes within the space domain of the size of  $\sim 1/p_T$ . 
Thus,  we can expect the limiting behaviour when the 
soft particle wave-length, 
$1/p_T$, becomes comparable with the  coherence range, at most 
with the nuclear radius $r_A$ (for the central
$AA$-collisions) or with the size of the region, where two nuclei overlap,
that is, for instance, for the $AuAu$ collisions $p_T< 1/r_A\sim 30$ MeV.
Inspection of the RHIC data \cite{star}~-~\cite{phobos}
shows, that, indeed, the ratio
\begin{equation}
R_{AB}^{N_{coll}}=\frac{1}{N_{coll}}\frac{dN_{AB}/dp_T}{dN_{pp}/dp_T},
\label{RAB}
\end{equation}
decreases with decreasing transverse momentum  $p_T$ of the secondaries. 
The
ratio $R_{AB}^{N_{coll}}$ (\ref{RAB}) allows to compare 
the particle density for 
the nucleus-nucleus collisions 
with that in the proton-proton interactions times the mean number of
nucleon-nucleon collisions $N_{coll}$, which is 
calculated using the Glauber model. 
Another way used
to present the RHIC data is to replace the number of
collisions $N_{coll}$ by the number of nucleons   
participating in the
interaction, $N_{part}$.
Since in the $pp$-collisions just two protons scatter, the
normalization factor is $N_{part}/2$ rather than $N_{coll}$.

Similar to the  limiting distributions, 
%discussed above,
corresponding to the soft bremsstrahlung
in the $e^+e^-$ and $pp$ processes,
we, first of all, expect
% at first
the energy-independent behaviour of
the soft particle densities. The existing data confirm,
that at low transverse momenta, $p_T<0.3$ GeV, the energy dependence of the ratios
$R_{AB}$ is, indeed, quite flat.
At so low $p_T$'s the ratios $R_{AB}$, measured 
by PHOBOS \cite{phobos}
at $\sqrt{s}=62.4$ and 200 GeV, largely coincide for all studied centrality ranges
(see Fig.~8 and Fig.~32 in \cite{phobos} for the normalizations with $N_{coll}$
and  $N_{part}$ respectively). 
At  lower energies
$\sqrt{s_{NN}}=5$ and 17~GeV the results were obtained by PHENIX 
\cite{phenix} (see Fig.~47), and a comparison is possible for  $p_T> 0.4$
GeV. A convergence of the $p_T$ spectra towards low $p_T$ 
at the two energies can be seen, although not yet a coincidence in the
observed  $p_T$-range. 
%for comparison  with $pp$ at different energies - Fig.26 of PHENIX.\\
%

Let us now turn to the limiting density 
of the soft particle production.
The data at the  lowest $p_T\approx 0.03$ GeV were collected by 
PHOBOS~\cite{phobos} (Fig. 3), the data with $p_T>0.2$ GeV 
were obtained by PHENIX 
\cite{phenix} (Fig. 11) and by STAR \cite{star1} (Fig. 46, Tab. 33) 
in the Au+Au collisions at 200 GeV. If we extrapolate the invariant
pion spectra from PHENIX and STAR by the exponential form at $p_T<0.5$ GeV 
we arrive at 
$I_0^{AuAu}\approx 1270$ GeV$^{-2}$ to be compared with 
 $I_0^{pp}\approx 7.8$ GeV$^{-2}$, found with the same procedure
for the  $pp$ collisions \cite{star1}.
However, the PHOBOS data at very low $p_T$ indicate a flattening of the
distribution according to the functional form (\ref{thermal}) with 
$T=0.229$ GeV. 

Fitting the PHOBOS and STAR $AuAu$ data for pions at $p_T<0.5$ GeV using
this parametrization, we obtain for the mean pion density at $p_T=0$
in the average
\begin{equation}
\text{central}\ AuAu\ \text{collisions (pions)}:\qquad I_0\simeq (950\pm 100)\ \text{GeV}^{-2},  
\label{I0auau}
\end{equation} 
which is about 30\% lower than the result of exponential parametrization.   
Assuming a similar fit for the low $p_T$ $pp$ STAR data, we
arrive at $I_0^{pp}\simeq 5.9$ GeV$^{-2}$ with $T=0.182$ GeV, therefore, the limiting density for the
nuclear collisions is
%
% 
%We first extrapolate the invariant distributions over $p_T$
%to the limit  $p_T=0$
%In a first look at the numbers we extrapolate the $p_T$ spectra of 
%invariant densities to $p_T=0$ 
%and after averaging arrive
% for the mean charged particle density at
%$I_0\simeq 830$ GeV$^{-2}$, 
% It may be interesting to compare the particle
%density observed by PHOBOS (see also the fit in the caption to Fig.3) and
%PHENIX with the proton-proton data presented in Table1.\\
%
%From Fig.3(PHOBOS) and Fig,11 (PHENIX) we have  in the central $AuAu$ collisions
%$I_0\simeq 700$ GeV$^{-2}$, 
more than 100 times larger than the corresponding value
in the $pp$ interactions. % in Tab.  1.  
On the other hand, in this case the Glauber model
calculation~\cite{phobos} gives $N_{coll}=1040$ and $N_{part}/2=172$
($\pm15\%$). 
Clearly, the soft particle density is about 10 times lower 
than that expected in
the case of independent (incoherent) collisions.
%ith the same reduction of the intensity of the soft particles. 
Indeed, from the extrapolation 
%reported above 
of the $p_T$ spectra
in the $AuAu$ and $pp$
collisions by either an exponential or the  thermal function 
(\ref{thermal}), as presented above,
 we find the ratio
\begin{equation} 
I_0^{AA}/I_0^{pp}\approx 160\pm 17, 
\label{AA/pp}
\end{equation}
which agrees 
with the calculated $N_{part}/2$, and, therefore, 
\begin{equation}
p_T\to 0: \qquad   R_{AA}^{N_{part}}\to 1\quad \text{and}\quad 
I_0^{AuAu} \approx  \dfrac{N_{part}}{2}\ I_0^{pp}. 
\label{RAApart}
\end{equation}  
Here $ R_{AA}^{N_{part}}$ is defined as in (\ref{RAB}), but with 
$N_{part}/2$ as a normalizing quantity.

A detail experimental
study was performed by PHOBOS \cite{phobos} (Fig. 32) for different 
ranges of  centrality, that is, the number of participants, at the two
energies 62.4 GeV and 200 GeV. Remarkably, the ratio $R_{AA}^{N_{part}}$ 
%defined as in (\ref{RAB}) but with $N_{part}/2$ as normalizing quantity,
approaches about unity
 in the soft limit for all selections of centralities
or the $N_{part}$ parameter. STAR \cite{star2}  measured the ratio
$R_{AA}^{coll}$ with high precision for $p_T>0.5$ GeV. It is found that this quantity
decreases below $p_T<2$ GeV and the extrapolation below the
measured values of  $p_T$
suggests the limiting behaviour (\ref{RAApart}) as well.

That is, the data are
consistent with the model where each pair of participating nucleons produces
its own number of secondaries independently of the {\it number of collisions}
each nucleon is participated in. This is related to the idea, that the overall
characteristics of bulk particle production 
depend only on the number of `wounded
nucleons' \cite{bialas} and not on the number of rescatterings.

\subsection{Antenna pattern in nuclear collisions}
Within the bremsstrahlung scenario this result suggests that there is a coherent
particle production over the range of nucleon size, but different nucleons
are separated in space-time so that their contributions remain incoherent
\footnote{The coherence is destroyed by the presence of different recoil
nucleons.}.  This is in agreement with the observation that no specific
structure reveals itself in the nuclear collisions at very small $p_T\sim
30$ MeV corresponding to the nuclear size (see, for example, Fig.  3 in
\cite{phobos} with pions in this range).

\begin{figure}[h]
\begin{center}
\epsfig{file=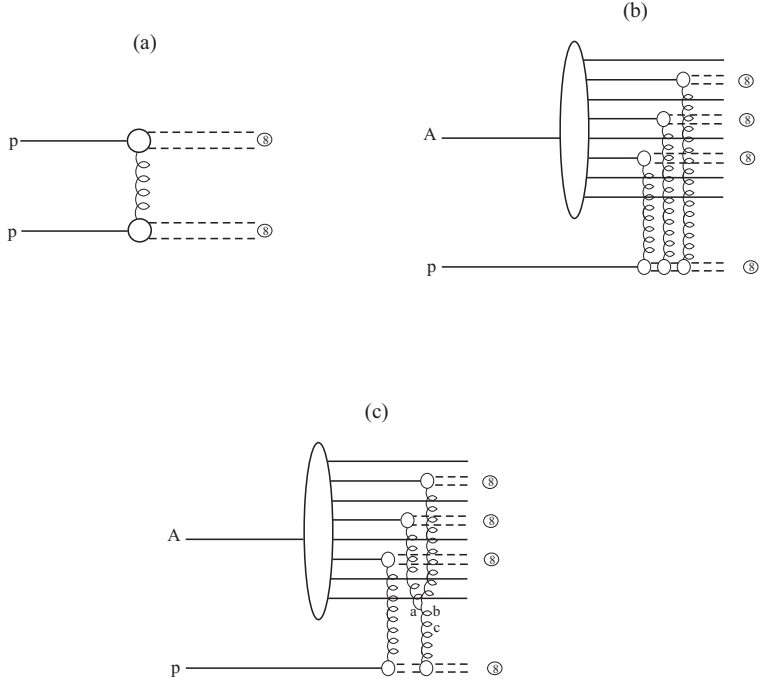,width=10.cm}
\end{center}
\caption{Diagrams, contributing to the $pp$ and $pA$ collisions in the minimal model
for soft particle production: (a) In the $pp$ collisions the exchanged gluon 
interacts with the colour triplet constituents $q$ or $qq$ in the proton to form an
outgoing colour octet system; (b,c) In the $pA$ collisions the proton can rescatter
inside the nucleus and then forms a colour octet system again. This implies that
the multiple gluon exchange acts as a single gluon exchange in the particle
production. In this example: $N_{coll}=3,\ N_{part}/2=2$.}
\label{fig:AA}
\end{figure}

 In spite of the multiple re-interactions of the
same nucleon in the nucleus which proceed through the corresponding number
of gluon exchanges in the  minimal model for the $pp$
collisions considered above, we still expect the limiting behaviour.  
The low $p_T$ particles
emitted in this process interact coherently with all t-channel gluons and,
actually, probe the overall
colour flow originated by this (t-channel) system of gluons. On the other hand, 
as we discussed in Sect. 3.1, 
even a large number of gluons produces dominantly the same octet colour flow
as in the $pp$ inelastic collision 
 (first, the octet exchange gives the dominant Leading Logarithmic
contribution, both in the DGLAP and in the BFKL kinematics, next -
interacting with the individual quark or with the colourless $q\bar q$- or
'quark-diquark'-system we can transfer only the colour octet quantum number).  
Examples of diagrams contributing in this minimal model to the
production of the low $p_T$ particles in the  $pp$ and $pA$ collisions are shown in Fig. 
\ref{fig:AA}. In the $AA$ collisions re-interactions of nucleons may occur as in the
$pA$
collisions. In these depicted processes the colour flow and the associated  
soft particle production are determined by the number of participating
nucleons irrespectively of the number of rescatterings, in agreement with the
phenomenological result (\ref{RAApart}).

\subsection{Universality of particle ratios at low $p_T$}
\begin{figure}[h]
\begin{center}
\begin{minipage}[h]{0.5cm} %0.1\textwidth}
\vspace{-4.5cm}
%{\bf
%{$ K^-/\pi^-$}\\
%{\rm ratio}
%}
\begin{displaymath} 
{ \frac{{\mbf K^-}}{{\mbf \pi^-}}}%  \ {\rm ratio}
\end{displaymath}
\end{minipage}
\begin{minipage}[h]{10cm}%0.1\textwidth}
%\begin{center}
\epsfig{file=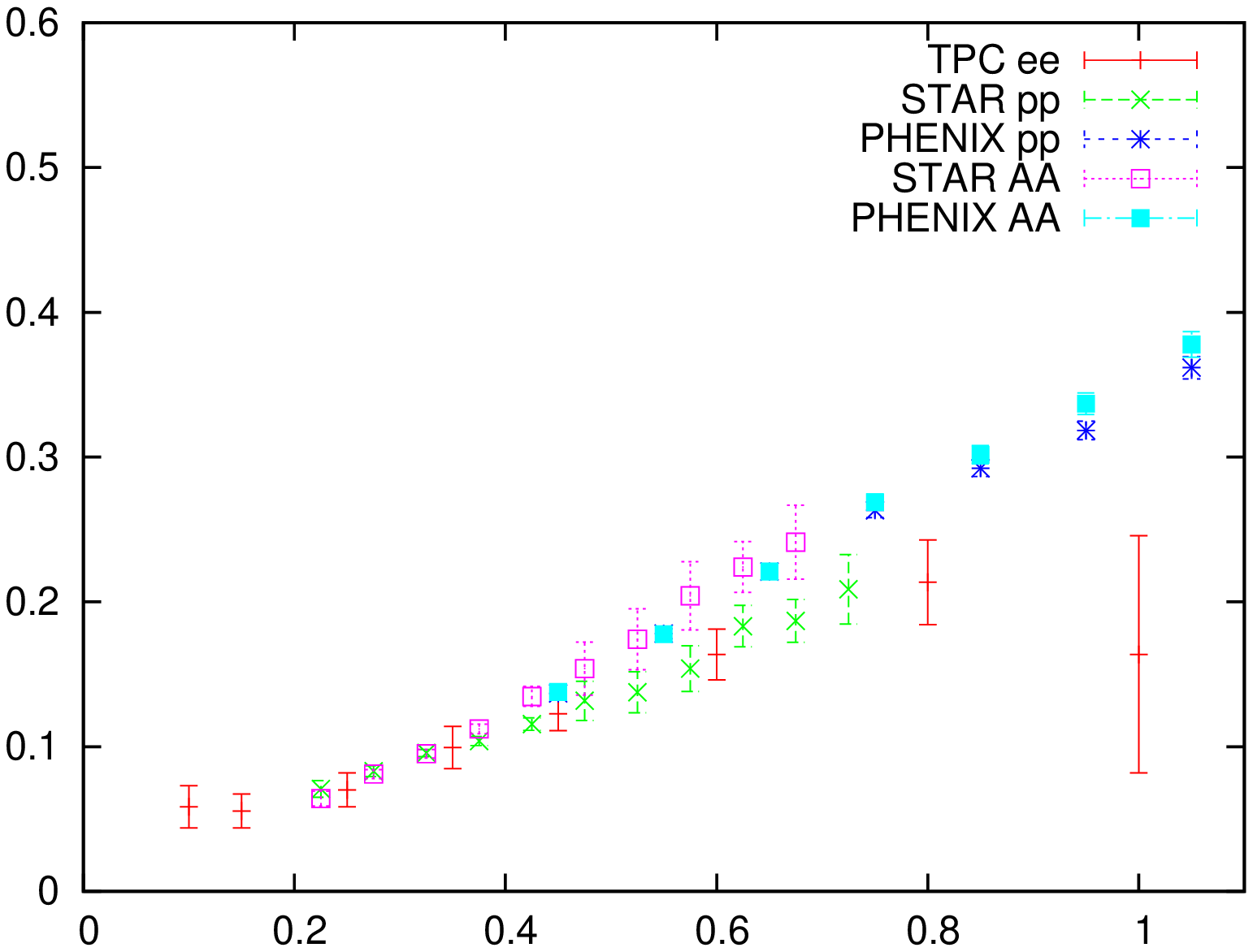,angle=-90,width=10.cm}
\hspace*{7cm} {\bf ${\bf p_T}$ [GeV]}
\end{minipage}
\end{center}
\caption{Convergence of particle ratios $K^-/\pi^-$ towards small $p_T$ for 
various processes: $e^+e^-$ annihilation 
(TPC \cite{tpc} data, $p_T$ with respect to sphericity axis), 
$pp$ (minimum bias) and central (0-5\%) $AuAu$
collisions (STAR \cite{star1} and PHENIX \cite{phenix1}
Collaborations).}
\label{fig:ratio}
\end{figure}
Having in mind this universal mechanism for soft particle production we may
ask whether there are any consequences for the relative rates of different
hadron yields in the same limit.  If the particles at low momenta in the
hadronic and nuclear collisions are related to the universal bremsstrahlung
from the (incoherent) superposition of primary colour octet charges, then,
in this limit, the particle composition should be the same in the different
processes.  To the extent, that the dominant underlying mechanism in the
$pp$ collisions is the gluon exchange between the quark constituents of the
proton with the initial and final bremsstrahlung, the $e^+e^-$ and hadronic
data on the particle ratios should approach each other as well.

A similarity of particle ratios $K/\pi$ and $\bar p/\pi$ in the
$e^+e^-$ and $pp$ reactions at $p_T<0.5 $ GeV
has been indeed noted already some time ago by the TPC collaboration
\cite{tpc}. In this measurement the transverse momentum $p_T$ for the particles in $e^+e^-$
collisions was defined with respect to the sphericity axis.

The $p_T$ dependence of the particle ratios for several hadronic collisions
have been compared by PHENIX \cite{phenix1}.  While at the large $p_T>2$ GeV
the ratios $p/\pi$ and $K/\pi$ tend to approach large values $\sim 1$ in the
central $AuAu$ collisions, these ratios are reduced for non-central and
minimum bias $pp$ collisions.  Remarkably, these ratios converge for the
different processes towards lower $p_T<1$ GeV.  In Fig.  \ref{fig:ratio} we
collect data in the low $p_T$ region on the ratio $K^-/\pi^-$ from the
$e^+e^-$, $pp$ and $AA$ interactions.  As one can see, the particle ratios,
indeed, approach each other towards low $p_T<0.4$ GeV pointing towards a dominance
of multiple $q\bar q$ dipole radiation in all processes.

\section{Conclusions}
An issue of  universality of  particle production in  various 
collision processes predated QCD. 
The perturbative dynamics based on the
QCD gluon bremsstrahlung
suggests a particular type of universality, which, however,
allows only for the predictions for the soft limit $p_T\to 0$.
This is the case,
when the lowest
order Born term with  its energy independence and elementary colour factors 
dominates,
whereas for the larger $p_T$'s the (non-universal) higher order contributions
take over. 
The predictions from this approach are confirmed with high accuracy in the
$e^+e^-$ measurements, where the energy independence of the soft particle
yield is observed,  and its intensity in quark and gluon jets 
is in agreement with the expectation from corresponding colour factors 
$C_F$ and $C_A$.

In this paper we discuss the production of low momentum particles in hadronic
collisions. Surprisingly, the soft particle production density in the $pp$ 
collisions is practically constant over the large energy range up to 1800 GeV.
Moreover, the approximate energy independence holds also in the
nuclear collisions at RHIC energies.
The colour factors related to the primary interaction process
reveal themselves also in the 
comparison between  the $e^+e^-$ and $pp$ collisions, 
according to a minimal model with gluon exchange in the
case of  hadronic processes.
Such a  comparison  would strongly benefit
from more precise measurements
of the low $p_T$ particle spectra in $pp$ collisions,
which could reveal the expected deviations from the exponential behaviour.
In the nuclear collisions the perturbative expectations,
based on the coherent  bremsstrahlung in a multi-gluon exchange 
process lead to %\mpar{***}
%concern 
the energy independence of the soft particle yield.
Moreover, the % \mpar{***}
% as well as its 
magnitude should scale with the number of
participant nucleons, $N_{part}$. 

As a result, the yield of the low $p_T$ particles can be directly related to the
soft QCD gluon bremsstrahlung from the colour charges, which are created in the
primary hard or semi-hard interaction, and is proportional to $C_F,\ C_A$ and
$\frac{N_{part}}{2}C_A$ for $e^+e^-,\ pp$ and $AA$ collisions respectively.  This concept
of universality is also supported by the observed convergence of the
$\pi:K:p$ particle ratios in all these processes in the $p_T\to 0$ limit.
 
From the first sight, this seems
to contradict the idea of thermalisation 
in the particle production.
However, we recall, that according to the space-time picture of particle
production, the central soft hadrons in the nucleon-nucleon 
$c.m.s.$ system are those,
which are formed first, and these soft particles with their large wave-length
probe only the colour charges participating in the primary local interaction
%probe only the primary  and outgoing colour charges 
%participating in the local interaction
at a transverse size of $\sim 1$ f.
The fast particles are formed later on, and leave the interaction region
without re-scattering on the slow hadrons which stay in the interaction
region. Therefore, the slow particles
may be not in a thermo-equilibrium with the whole
hadronic system.
At the very early stages following  the primary interaction no
re-interactions of produced hadrons from the different nucleons will take
place, while the re-interactions of one
nucleon with other nucleons in the nucleus should not
be accounted for because of the coherent nature of soft particle emission.
Then, the universality of this soft particle
production from the primary colour sources becomes a plausible 
phenomenon, since it basically reflects the
universality of soft particle production from the individual isolated
quark-antiquark dipoles.

It will be very interesting to extend
such measurements on the limiting soft particle production in 
$pp$ and $AA$ collisions 
to the higher energies at the LHC  and to see whether 
any new incoherent sources appear. Such measurements could set a critical
benchmark for the models of 
multiparticle production in hadronic interactions. 

\section*{Acknowledgements}
We thank Gosta Gustafson, Frank Krauss,  Risto Orava, Peter Seyboth and Torbjorn Sjostrand
for useful discussions.
VAK thanks the Theory Group of
the Max Planck Institute for hospitality.
This work was supported by the grant RFBR
10-02-00040-a, by the Federal Program of the Russian State RSGSS-3628.2008.2.

\end{document}